\documentclass[12pt]{article}
\usepackage[body={17cm, 23cm, centered}]{geometry}

\usepackage{lmodern}
\usepackage{epsfig,verbatim,cite}
\usepackage{braket}

\usepackage[T1]{fontenc}
\usepackage[utf8]{inputenc}
\usepackage[unicode]{hyperref}
\hypersetup{colorlinks=true,linktocpage,breaklinks,
            urlcolor=blue,
            linkcolor=red,
            citecolor=blue
            }

\usepackage{float} 
\usepackage{subcaption}

\usepackage{longtable}

\usepackage{mathtext,marvosym,textcomp}
\usepackage{mathtools}
\usepackage{slashed}
\usepackage[dvipsnames,svgnames,table]{xcolor}

\usepackage{tikz}
\usetikzlibrary{decorations.pathmorphing}
\tikzset{snake it/.style={decorate, decoration=snake}}
\usetikzlibrary{positioning}
\usetikzlibrary{calc}

\usepackage[ normalem]{ulem}
\usepackage{tocloft}
\usepackage{epsfig,amsmath,amsfonts,amssymb,arydshln}
\usepackage[makeroom]{cancel}

\def\a{\alpha} 
\def\b{\beta} 
\def\g{\gamma} 
\def\d{\delta}

\def\h{\eta} 

\def\k{\kappa} 
\def\l{\lambda} 
\def\m{\mu}

\def\ba{\bar{a}}

\def\fr{\frac}  

\def\ph{\phantom}

\def\mc{\mathcal}

\def\tf{\tilde{f}}

\def\tT{\tilde{T}}

\def\tf{\tilde{f}}

\def\rmGL{\mathrm{GL}}
\def\rmU{\mathrm{U}}
\def\rmO{\mathrm{O}}

\def\frg{\mathfrak{g}}

\def\tfrg{\tilde{\mathfrak{g}}}

\def\ba{\begin{array}{r@{}l@{}} }
\def\ea{\end{array}}

\def\baa{\begin{array}{r@{}l  r@{} l@{} } }
\def\ea{\end{array}}

\def\rk{\mathrm{rank}\,\k}

\newcommand{\algdot}[2]{
\filldraw[black] (#1,#2) circle (0.05);
}

\newcommand{\algtwo}[3]{
    \draw[<->, thick,  rotate around={#3:(#1,#2)}] (#1-0.75,#2) -- (#1,#2);
    \draw[<->, thick,  rotate around={#3:(#1,#2)}] (#1,#2) -- (#1+0.75,#2);
    \draw[decorate, decoration={snake, segment length=4, amplitude=1}, thick, rotate around={#3:(#1,#2)}] (#1,#2-0.2) -- (#1,#2+0.2);
     \node at ({#1 + 0.3*sin(#3)},{#2-0.3*cos(#3)}) (1) {\footnotesize{$0$}};
     }

\newcommand{\algfive}[2]{
    \draw[-, thick] (#1,#2) -- (#1-0.7,#2-0.7) -- (#1+0.7,#2-0.7) -- (#1+0.7,#2+0.7) -- (#1,#2);
    \draw[decorate, decoration={snake, segment length=4, amplitude=1}, thick] (#1,#2) -- (#1+0.7,#2);
    \draw[decorate, decoration={snake, segment length=4, amplitude=1}, thick] (#1,#2) -- (#1,#2-0.7);
    \node at (#1+0.8,#2+0.8) (1) {\footnotesize{$\gamma$}};
    \node at (#1+0.8,#2-0.7) (1) {\footnotesize{$\beta$}};
    \node[rotate=45] at (#1-0.1,#2+0.1) (1) {\tiny{$(0,0)$}};
}

\newcommand{\algsix}[2]{
    \draw[-, thick] (#1,#2-0.5) -- (#1-0.7,#2-0.5) -- (#1-0.7,#2+0.5) -- (#1+0.7,#2+0.5) -- (#1+0.7,#2-0.5) -- (#1,#2-0.5);
    \draw[decorate, decoration={snake, segment length=4, amplitude=1}, thick] (#1,#2-0.5) -- (#1,#2+0.5);
    \node[rotate=90] at (#1-0.85,#2) (1) {\footnotesize{$-\infty$}};
    \node[rotate=90] at (#1+0.85,#2) (1) {\footnotesize{$+\infty$}};
    \node at (#1,#2+0.65) (1) {\footnotesize{$\infty$}};
    \node at (#1,#2-0.65) (1) {\footnotesize{$0$}};
    \node at (#1+0.8,#2-0.5) (1) {\tiny{$\a$}};
    \node at (#1+0.8,#2+0.65) (1) {\tiny{$\beta$}};

}

\newcommand{\algeight}[3]{
    \draw[-, thick, rotate around={#3:(#1,#2)}] (#1-0.5, #2) -- (#1+0.5, #2);
    \filldraw[black, rotate around={#3:(#1,#2)}] (#1-0.5,#2) circle (0.05);
    \filldraw[black, rotate around={#3:(#1,#2)}] (#1+0.5,#2) circle (0.05); 
    \node at ({#1+0.7*cos(#3)},{#2+0.7*sin(#3)}) (1) {\footnotesize{$\b$}}; 
}

\newcommand{\algnine}[3]{
    \draw[->, thick, rotate around={#3:(#1,#2)}] (#1-0.5, #2) -- (#1+0.5, #2);
    \filldraw[black, rotate around={#3:(#1,#2)}] (#1-0.5,#2) circle (0.05);
    \node at ({#1+0.7*cos(#3)},{#2+0.7*sin(#3)}) (1) {\footnotesize{$\a$}}; 
    }

\begin{document}

\renewcommand{\contentsname}{}
\renewcommand{\refname}{\begin{center}References\end{center}}
\renewcommand{\abstractname}{\begin{center}\footnotesize{\bf Abstract}\end{center}} 

\begin{titlepage}
\ph{preprint}

\vfill

\begin{center}
   \baselineskip=16pt
   {\large \bf O(4,4) dualities and Manin triples
   }
   \vskip 1cm
    Angelina Kurenkova$^{a}$\footnote{\tt askurenkova@mail.ru }
    Edvard T. Musaev$^{b,c}$\footnote{\tt emusaev@theor.jinr.ru},
       \vskip .6cm
             \begin{small}
                          {\it                           
                          $^a$Institute of Theoretical and Mathematical Physics, Moscow State University, 119991, Russia\\
                          $^b$Bogoliubov Laboratory of Theoretical Physics, Joint Institute for Nuclear Research, 6, Joliot Curie, 141980 Dubna, Russia\\
                          $^c$Moscow Institute of Physics and Technology, 
                         141702, Dolgoprudny, Russia
                          } \\ 
\end{small}
\end{center}

\vfill 
\begin{center} 
 {Abstract}
\end{center} 
\begin{quote}
We provide a coarse classification of all 8-dimensional Manin triples, that describe Poisson--Lie T-dualities between 4-dimensional group manifold solutions to supergravity equations. We find several such dualities and one Poisson--Lie triality. For each class we refine the classification by marking whether the corresponding Drinfeld double can be obtained by a Yang--Baxter deformation of the initial algebra paired by the four-dimensional Abelian algebra.
\end{quote}

\vfill
\setcounter{footnote}{0}
\end{titlepage}

\tableofcontents
\setcounter{page}{1}

\section{Introduction}

Moduli space of string vacua possesses rich algebraic structure due to the high amount of string symmetries referred to as dualities. Relevant for the analysis presented here are T-duality symmetries, that have first been observed as symmetries of the string partition function in \cite{Buscher:1987qj} and of string background field equations in \cite{Buscher:1987sk}. In modern classification these are referred to as abelian T-dualities and are such transformations of the string background given by the metric, Kalb--Ramond field, the dilaton and R-R gauge fields, that leave the partition function invariant. At the level of supergravity these are solution generating transformations and can be used e.g. to generate backgrounds with non-geometric fluxes (T-folds \cite{Hull:2004in,Hull:2006qs}).

Abelian T-dualities relate to toroidal backgrounds and map two $\rmU(1)^d$ isometry groups, where $d$ stands for dimensions of the torus. A generalization to non-abelian isometries has been found in \cite{delaOssa:1992vci}. Such non-abelian T-dualities (NATD) are also solution generating transformations, however, the initial (non-abelian) symmetry gets broken upon dualisation, raising the question of finding the inverse. The corresponding procedure has been formulated in \cite{Klimcik:1995ux,Klimcik:1995dy} for 2d sigma-models in terms of algebraic structures known as the classical Drinfeld double. The idea is that for a duality to exist between two backgrounds it must be possible to construct a classical Drinfeld double in terms of Manin triple $(\frg,\tfrg,\h)$. Here $\frg$ denotes algebra of the initial symmetries $\tfrg$ is the so-called dual algebra (more detail below) and the quadratic form $\h$ defines isotropic structure on the algebra $\h(T_a,\tT^b)=\d_a{}^b$, where $T_a$ and $\tT^a$ stand for basis of $\frg$ and $\tfrg$ respectively. The isometry algebra $\frg$ is defined by its structure constants as usual
\begin{equation}
    [T_a,T_b] = f_{ab}{}^cT_c.
\end{equation}
Structure constants $\tf_a{}^{bc}$ of the dual algebra reflect non-conservation of the standard N\"other currents
\begin{equation}
    dJ^a = \tf_a{}^{bc}J_b\wedge J_c.
\end{equation}
Note that since the currents do not conserve in the standard sense, the initial isometries are hidden, that is precisely how symmetries are realized in a non-abelian T-dual background. 

The overall consistency of the construction can be summarized in the following algebraic equations
\begin{equation}
\label{eq:DDcond}
    \begin{aligned}
        2 f_{e[a}{}^c\tf_{b]}{}^{ed} - 2 f_{e[a}{}^d\tf_{b]}{}^{cd} = f_{ab}{}^e\tf_{e}^{cd}.
    \end{aligned}
\end{equation}
These are nothing but the consistency conditions for Drinfeld double defined by commutation relations
\begin{equation}
    \begin{aligned}\relax
        [T_a,T_b] &= f_{ab}{}^cT_c, \\
        [\tT^a,\tT^b] &= \tf_{c}{}^{ab}\tT^c, \\
        [T_a,\tT^b] & = \tf_a{}^{bc}T_c - f_{ac}{}^b \tT^c.
    \end{aligned}
\end{equation}
The quadratic form $\h$ induces an $\rmO(d,d)$ structure on the algebra, that naturally corresponds to the $\rmO(d,d)$ abelian T-duality symmetry. A generalization to U-duality symmetries has been developed in the series of works \cite{Sakatani:2019zrs,Malek:2019xrf,Blair:2020ndg} where the notion of exceptional Drinfeld algebra (EDA) has been formulated. Examples of non-abelian U-dualities of 11d backgrounds have been presented e.g. in \cite{Musaev:2020nrt,Blair:2022gsx}.

Certainly, the conditions \eqref{eq:DDcond} will fail to satisfy for a randomly chosen pair of algebras, hence a list of possible Manin triples is required to construct pairs of dual solutions with the corresponding (hidden) symmetries. Since $\frg$ is a real Lie algebra it is natural to start with known classifications of real Lie algebras and list all possible duals in each class. This idea stands behind the classifications \cite{Snobl:2002kq} of real 6d Manin triples, \cite{Hlavaty:2020pfj} of 6d EDA's, \cite{Kumar:2023uyu} of 10d EDA's and of the present paper for 8d Manin triples. During preparation of the paper we have learned of the work \cite{Hlavaty:2025zfh} appeared recently, that presents very similar results based, however, on a different classification. We give more details on the differences of the approaches and the results in the next section.

The paper structured as follows. Section \ref{sec:main}  contains main results of the paper, that is the set of dualities between 4d supergravity backgrounds presented as a set of Manin triples with both $\frg$ and $\tfrg$ indecomposable (self dualities are also not included). In Section \ref{sec:detail} we provide a detailed description of our approach to the classification. Section \ref{sec:list} contains all results listed as duality classes for each initial isometry algebra $\frg$.

\section{Main results}
\label{sec:main}

Classification of 8d Manin triples presented here is based on the classification of 4d real Lie algebras by Mubarakzyanov \cite{mubar} (see \cite{Popovych:2003xb} for the English version). Table \ref{tab:list4d} contains all real 4d Lie algebras listed according to this classification. Since our classification of Manin triples is based on the same principle as the classification by Mubarakzyanov, it is worth giving more details on how the algebras are split into classes in Table \ref{tab:list4d}.

\begin{longtable}{ | r | l || r | l || r | l |}

    \hline
    $\frg_{4,1}$ & $ \ba [T_2, T_4]&{}= T_1 \\{} [T_3, T_4] &{}= T_2 \ea$  & $\frg_{4,5}$ &$\ba [T_1,T_4]&{}=T_1\\{} [T_2,T_4]&{} = \b T_2\\{} [T_3,T_4]&{} =\g T_3\\ \b\g {}& \, \neq 0 \\ -1 \leq \g  &\,\leq \b \leq 1\ea$ &  $\frg_{4,9}$& $\ba [T_2,T_3]&{}=T_1 \\{} [T_1,T_4]&{}=2 \a T_1 \\{} [T_2,T_4]&{}=\a T_2 - T_3 \\{} [T_3,T_4]&{}=T_2 + \a T_3 \\{} \a&{} \geq 0 \ea$ \\
    \hline
    $\frg_{4,2}$& $\ba [T_1, T_4]&{}=\beta T_1 (\beta \neq 0) \\{} [T_2, T_4]&{}=T_2 \\{} [T_3,T_4]&{}=T_2+T_3\ea$ & $\frg_{4,6}$ &$ \ba [T_1,T_4]&{}=\a T_1 \\{} [T_2,T_4]&{} = \b T_2 - T_3 \\{} [T_3,T_4] &{} = T_2 + \b T_3\\ \a&{} \neq 0, \b\geq 0 \ea$ & $\frg_{4,10}$&$\ba [T_1,T_3]&{}=T_1 \\{} [T_2, T_3]&{}=T_2 \\{}[T_1,T_4]&{}= -T_2 \\{} [T_2,T_4]&{}=T_1 \ea$ \\ 
    \hline
    $\frg_{4,3}$& $\ba [T_1,T_4]&{}=T_1 \\{} [T_3,T_4]&{}=T_2 \ea $ & $\frg_{4,7}$ & $\ba [T_2,T_3]&{} =T_1 \\{} [T_1,T_4]&{} =2T_1\\{}[T_2,T_4]&{} =T_2 \\{} [T_3,T_4]&{}=T_2 + T_3 \ea$ & &   \\
    \hline
    $\frg_{4,4}$& $\ba [T_1,T_4]&{}=T_1\\{} [T_2,T_4]&{}=T_1 + T_2 \\{} [T_3,T_4]&{}=T_2 + T_3\ea $ &  $\frg_{4,8}$& $\ba [T_2,T_3]&{} =T_1 \\{} [T_1,T_4]&{} =(1+\beta)T_1\\{} [T_2,T_4]&{}=T_2 \\{} [T_3,T_4]&{}=\beta T_3  \\{} \beta &{} \in [-1,1]\ea$ & & \\ 
    \hline 
\caption{Classification of 4-dimensional indecomposable real Lie algebras $\frg_{4,n}$ with $n=1,\dots,10$. }
\label{tab:list4d}
\end{longtable} 

The central object is the maximal nilpotent ideal $\mc{I}$ of the algebra. If dim$\mc{I}=4$ there is only one indecomposable algebra that is $\frg_{4,1}$. If dim$\mc{I}=3$ one has two options $\mc{I}=3\frg_1$ and $\mc{I}=\frg_{3,1}$ (Bianchi II). In the first case the algebra can be brought to the form 
\begin{equation}
    [T_\a,T_*] = M_\a{}^\b T_\b,
\end{equation}
where $\a,\b=1,2,3$ and $T_*$ is the fourth generator. Algebras in this set are classified by the canonical forms of the matrix $M$, that gives $\frg_{4,2},\dots,\frg_{4,6}$. 

In the second case, when $\mc{I}=\frg_{3,1}$ the basis can be chosen such that
\begin{equation}
    [T_2,T_3]=T_1,
\end{equation}
and the algebras are classified by canonical forms of the $(2,3)$ block of the matrix $M$. This gives $\frg_{4,7},\frg_{4,8}, \frg_{4,9}$. Finally, if dim$\mc{I}=2$ one has only one indecomposable algebra $\frg_{4,10}$.

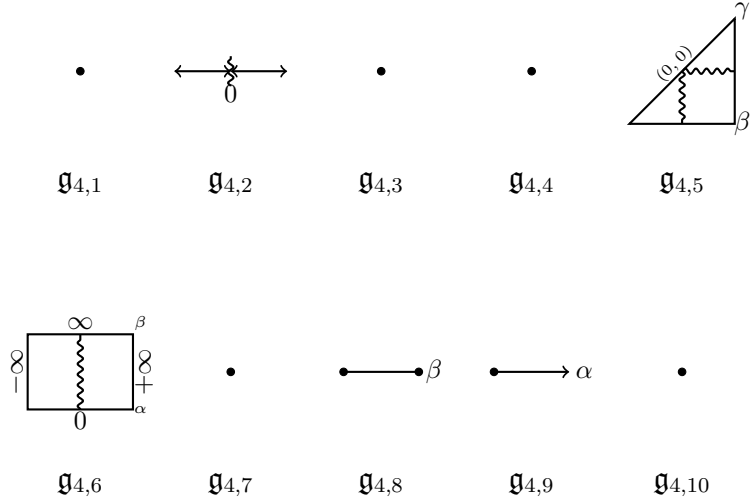
\begin{figure}[H]
    \centering  
    \begin{tikzpicture}
        \algdot{1}{1}  
        \node at (1,-0.5) (1) {$\frg_{4,1}$};
        \algtwo{3}{1}{0}
        \node at (3,-0.5) (1) {$\frg_{4,2}$};
        \algdot{5}{1}
        \node at (5,-0.5) (1) {$\frg_{4,3}$};
        \algdot{7}{1}
        \node at (7,-0.5) (1) {$\frg_{4,4}$};
        \algfive{9}{1}
        \node at (9,-0.5) (1) {$\frg_{4,5}$};
        \algsix{1}{-3}
        \node at (1,-4.5) (1) {$\frg_{4,6}$};
        \algdot{3}{-3}
        \node at (3,-4.5) (1) {$\frg_{4,7}$};
        \algeight{5}{-3}{0}
        \node at (5,-4.5) (1) {$\frg_{4,8}$};
        \algnine{7}{-3}{0}
        \node at (7,-4.5) (1) {$\frg_{4,9}$};
        \algdot{9}{-3}
        \node at (9,-4.5) (1) {$\frg_{4,10}$};
    \end{tikzpicture}
    \caption{Schematic depiction of four-dimensional Lie algebras reflecting moduli space of each class. Wave line means that the parameter cannot take values there, arrow means that the parameter can go to infinity. Algebras with no free parameters are depicted by a dot.}
    \label{fig:algebras}
\end{figure}

Since the dualities following from the classification presented in Section \ref{sec:list} work often for only particular sets of parameters, it is convenient to represent them diagrammatically as pictures. For that we depict each real 4d Lie algebra from the classification as a dot, line or triangle as shown in Fig.\ref{fig:algebras}.

Our classification algorithm works as follows. Using the Wolfram Mathematica file, that can be found in \cite{Kurenkova_8d_Manin_triples_2025}, we start with a 4d real Lie algebra $\frg_{4,n}$ and solve equations \eqref{eq:DDcond} together with Jacobi identities for $\tf_{a}{}^{bc}$ to find all possible sets of dual structure constants. Running the program results in the following table 
\begin{longtable}{|c|c|c|c|c|c|c|c|c|c|c|}
\hline
         Algebra              & $\frg_{4,1}$ & $\frg_{4,2}$& $\frg_{4,3}$& $\frg_{4,4}$& $\frg_{4,5}$& $\frg_{4,6}$& $\frg_{4,7}$& $\frg_{4,8}$& $\frg_{4,9}$& $\frg_{4,10}$  \\
\hline
         \# of solutions & 9 & 3 & 7 & 2 & 4 & 4 & 3 & 4  & 4 & 21  \\
\hline
    \caption{Number of possible duals for each 4d real Lie algebra.}
    \label{tab:number_of_sols}
\end{longtable}
We call these possible duals since on one hand some of the are complex, thus reducing the number of classes, on other hand each solution in turn splits into various classes, thus increasing their number.

For each of the found solutions we repeat the analysis of Mubarakzyanov and split each found set of dual structure constants into cases depending on i) eigenvalues $\l_{1,2,3}$ of the Killing form $\k^{ab} = \tf_c{}^{ad}\tf_d{}^{bc} $, ii) whether the maximal nilpotent ideal $\mc{I}$ is $3\frg_1$ (abelian) or $\frg_{3,1}$ (Bianchi II), and iii) if $\mc{I}=3\frg_1$ the classes are distinguished by eigenvalues of the matrix $M_\a{}^\b$, which we call $\m_{1,2,3}$. Values of parameters of dual algebras (if any) are expressed in terms of eigenvalues, and hence are invariant under the change of the basis.

\begin{figure}[H]
    \centering
    \begin{tikzpicture}
        \algdot{2}{1}
        \algfive{0}{0}
        \algeight{4}{0}{0}
        \algdot{4}{0}
        \draw[<->, thick] (2.1,1) to[out=0, in=90] (4,0.1);
        \draw[<->, thick] (2,0.9) to[out=-90, in=45] (0.5,-0.5);
        \draw[<-, thick, red] (0,0) -- (0.7,-0.7);
        \node at (4,-0.2) (1) {\footnotesize{0}};
        \node at (2,1.3) (1) {\footnotesize{$\frg_{4,1}$}};
        \node at (4.5,-0.5) (1) {\footnotesize{$\frg_{4,8}$}};
        \node at (-0.5,0.5) (1) {\footnotesize{$\frg_{4,5}$}};
    \end{tikzpicture}\hspace{1cm}
    \begin{tikzpicture}
        \algtwo{0}{0}{0}
        \algfive{2}{0}
        \draw[<->, thick] (0.2,-0.5) to[out=-90, in=-90] (2,-0.8);
        \draw[<-, thick, red] (2,0) -- (2.7,-0.7) -- (1.3,-0.7);
        \node at (3.5,0) (1) {\footnotesize{$\frg_{4,5}$}};
        \node at (0,0.5) (1) {\footnotesize{$\frg_{4,2}$}};
    \end{tikzpicture}\\ \vspace{1cm}
    \begin{tikzpicture}
        \algdot{0}{0}
        \algeight{2}{0}{90}
        \draw[<->, thick] (0.1,-0.1) to[out=-40, in=-160] (1.9,-0.5);
        \node at (0,0.3) (1) {\footnotesize{$\frg_{4,3}$}};
        \node at (1.6,0) (1) {\footnotesize{$\frg_{4,8}$}};
        \node at (2.2,-0.7) (1) {\footnotesize{$-1$}};
    \end{tikzpicture}\hspace{1cm}
     \begin{tikzpicture}
        \algdot{0}{0}
        \algtwo{2}{0}{90}
        \draw[<->, thick] (0.1,-0.1) to[out=-40, in=-160] (1.9,-0.4);
        \node at (0,0.3) (1) {\footnotesize{$\frg_{4,4}$}};
        \node at (1.6,0.5) (1) {\footnotesize{$\frg_{4,2}$}};
        \node at (2.3,-0.4) (1) {\footnotesize{$-1$}};
        \algdot{2}{-0.4}
    \end{tikzpicture}\hspace{1cm}
     \begin{tikzpicture}
        \algsix{0}{0}
        \algnine{3}{0}{90}
        \draw[<->, thick] (0.3,-0.3) to[out=-15, in=-160] (2.8,0);
        \node at (0,-1) (1) {\footnotesize{$\frg_{4,6}$}};
        \draw[<->, thick, red] (0,-0.5) -- (0.7,0);
        \node at (2.6,0.5) (1) {\footnotesize{$\frg_{4,9}$}};
    \end{tikzpicture}\\ \vspace{1cm}
    \begin{tikzpicture}
        \algdot{0}{0}
        \algeight{3}{0}{90}
        \algtwo{1.5}{-2}{0}
        \algdot{1.1}{-2}
        \algdot{3}{-0.2}
        \node at (1,-2.32) (1) {\footnotesize{$-2$}};
        \node at (3.3,-0.2) (2) {\footnotesize{$-\fr12$}};
        \node at (0,0.2) (1) {\footnotesize{$\frg_{4,7}$}};
        \node at (2.6,0.4) (1) {\footnotesize{$\frg_{4,8}$}};
        \node at (2.2,-1.7) (1) {\footnotesize{$\frg_{4,2}$}};
        \draw[<->, thick] (0.1,-0.1) to[out=-15, in=-160] (2.9,-0.2);
        \draw[<->, thick] (0.1,-0.1) to[out=-15, in=90] (1.1,-1.9);
        \draw[<->, thick] (1.1,-1.9) to[out=90, in=-160] (2.9,-0.2);
    \end{tikzpicture}\hspace{1cm}
     \begin{tikzpicture}
        \algdot{0}{0}
        \algeight{2}{0}{90}
        \draw[<->, thick] (0.1,-0.1) to[out=-40, in=-160] (1.9,-0.5);
        \node at (0,0.3) (1) {\footnotesize{$\frg_{4,10}$}};
        \node at (1.6,0) (1) {\footnotesize{$\frg_{4,8}$}};
        \node at (2.2,-0.7) (1) {\footnotesize{$-1$}};
    \end{tikzpicture}
    \caption{Duality relations between 4d real Lie algebras: algebras on sides of an arrow can be part of a Manin triple. Red lines denote part of the algebras module space in a given class to be dualised. Algebras are not taken in the standard form. }
    \label{fig:duals}
\end{figure}
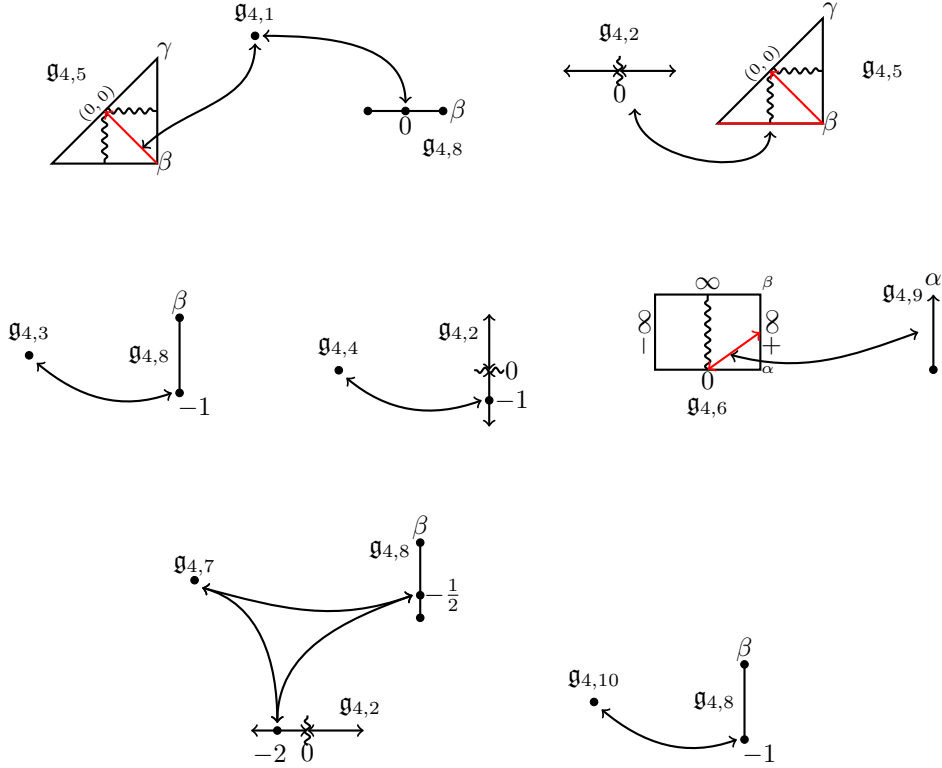

The last point is crucial, since our classification is coarse in the following sense. After solving the consistency conditions dual structure constants are not necessarily in the standard form as given by Table \ref{tab:list4d}. To bring them to the standard form one has to perform a $\rmGL(4)$ rotation \textbf{inside} the Drinfeld double, which will distort the form of the initial algebra $\frg$. Hence, one is interested in finding such a $\rmGL(4)$ transformation that preserves the standard form of $\frg$ and brings $\tfrg$ to the simplest possible form. Doing that one may find different classes of such transformations and further refine the classification. We are not doing this in the present paper, and hence in a given Manin triple, say $(\frg_{4,1},\frg_{4,8}(\b=0))$, the second algebra is not in the standard form. However, algebra parameters are invariant and stand as the are given in the list of Section \ref{sec:list}. Such refined classification for 6d Manin triples was presented in   \cite{Snobl:2002kq}. For this reason we prefer not to list dual commutation relations for each case, concentrating instead to duality relations between algebras. However, the Mathematica file \cite{Kurenkova_8d_Manin_triples_2025} can be used to generate an of the dual algebras in the list in the convenient form of commutation relations $[\tT^a,\tT^b]$.

For each of the 4d real indecomposable Lie algebras we find either themselves (in a different basis), a decomposable algebra or another 4d indecomposable Lie algebra. The latter case seems to be the most interesting and we depict all such pairs as arrows on diagrams in Fig. \ref{fig:duals}. Each pair of algebras sitting at two points of an arrow can form a Manin triple, with one of them taken in the standard form and another (in general) in a some other basis. Therefore, each arrow represent a Poisson--Lie T-duality between the corresponding group manifolds, given the proper geometric realization. We also find a PL-triality relation between $(\frg_{4,7},\frg_{4,8}^{\b=-1/2},\frg_{4,2}^{\b=-2})$, meaning that any pair of these three can form a Manin triple with one algebra taken in the standard form.

In addition to the coarse analysis we refine the classification as follows. It is known that given a Lie algebra $\frg$ a proper dual Lie algebra inside a Manin triple $(\frg, 4\frg_1)$ can be generated by a bi-vector Yang--Baxter deformation as
\begin{equation}
    \tf_c{}^{ab} = 2 r^{d[a}f_{cd}{}^{b]},
\end{equation}
where $r^{ab}$ satisfies classical Yang--Baxter equation. For each class we mark by [YB] those, that can be represented in this form. Remarkably, this is true for almost all classes found. This seems to be the most useful information presented here, since backgrounds corresponding to such double algebras can be easily generated by bi-vector deformations without the need of going through all the technicalities of the geometric realization procedure. Certainly, the full finest classification of real Manin triples should include both information about choice of the basis and about whether an family of triples is the result of a Yang--Baxter deformation.

\section{Detailed comments on the classification}
\label{sec:detail}

In this section we give a more detailed explanation of the algorithm that we follow to classify 8d Manin triples. As the concrete example we take the algebra $\frg_{4,1}$ whose first solution for possible duals represents all relevant issues. As the very step we take $f_{24}{}^1=1$, $f_{34}{}^2=1$ and solve the consistency conditions \eqref{eq:DDcond} together with the Jacobi condition for dual structure constants $\tf_a{}^{bc}$. Our Mathematica code gives a set of 9 different solutions. For the sake of a concrete example let us take the first one. The code gives the possible dual in the form of the following commutation relations
\begin{equation}
\begin{aligned}\relax
    [\tilde{T}^2, \tilde{T}^3]&=\overset{1}{g}_{4,2,3} \tilde{T}^4,\\
    [\tilde{T}^1, \tilde{T}^4]&=\overset{1}{g}_{2,1,2} \tilde{T}^4 ,\\
    [\tilde{T}^2, \tilde{T}^4]&=-\overset{1}{g}_{1,1,2} \tilde{T}^4 , \\
    [\tilde{T}^1, \tilde{T}^3]&=\overset{1}{g}_{4,1,3} \tilde{T}^4,\\
    [\tilde{T}^1, \tilde{T}^2]&=\overset{1}{g}_{1,1,2} \tilde{T}^1+\overset{1}{g}_{2,1,2} \tilde{T}^2+\overset{1}{g}_{3,1,2} \tilde{T}^3+\overset{1}{g}_{4,1,2} \tilde{T}^4,
\end{aligned}
\end{equation}
where $\overset{1}{g}_{a,b,c}$ correspond to $\tf_a{}^{bc}$ and the index $1$ denotes the number of a solution within the set of all possible duals. For $\frg_{4,1}$ the index can be $1,\dots,9$. To determine the maximal nilpotent ideal we first find the Killing form, that has the following matrix
\begin{equation}
    \k = 
        \begin{bmatrix}
 2 \overset{1}{g}_{2,1,2}^2 & -2 \overset{1}{g}_{1,1,2} \overset{1}{g}_{2,1,2} & 0 & 0 \\
 -2 \overset{1}{g}_{1,1,2} \overset{1}{g}_{2,1,2} & 2 \overset{1}{g}_{1,1,2}^2 & 0 & 0 \\
 0 & 0 & 0 & 0 \\
 0 & 0 & 0 & 0 \\
\end{bmatrix}.
\end{equation}
This is a rank one matrix and the eigenvalue is 
\begin{equation}
    \l=(\overset{1}{g}_{1,1,2})^2 + (\overset{1}{g}_{2,1,2})^2.
\end{equation}
Next we use Jordanian decomposition of the Killing form to find a suitable basis. Here one must distinguish cases with $\l\neq0$ and $\l=0$, as they correspond to different transformation matrices and Jordanian forms.

\textbf{1.}  $\l\neq 0$. In this case the Jordanian  form of $\k$ is diagonal
Since the only non-vanishing component is $\k_{4,4}$ it is now convenient to use Jordanian decomposition of the matrix $M_\a{}^\b$ to bring the commutation relations to the standard form. Note, that we do not need abelian maximal nilpotent ideal at this step, since we do not use canonical forms of $M$ for classification. The corresponding transformation matrix reads
\begin{equation}
    O=
        \begin{bmatrix}
 0 & 0 & -\frac{\overset{1}{g}_{3,1,2}}{\overset{1}{g}_{2,1,2}} & \frac{\overset{1}{g}_{3,1,2} \left(\overset{1}{g}_{2,1,2} \overset{1}{g}_{4,1,3}-\overset{1}{g}_{1,1,2} \overset{1}{g}_{4,2,3}\right)}{\overset{1}{g}_{2,1,2} \left(\overset{1}{g}_{1,1,2}^2+\overset{1}{g}_{2,1,2}^2\right)} \\
 \frac{\overset{1}{g}_{1,1,2}}{\overset{1}{g}_{2,1,2}} & 1 & \frac{\overset{1}{g}_{3,1,2}}{\overset{1}{g}_{2,1,2}} & \frac{\overset{1}{g}_{3,1,2} \left(\overset{1}{g}_{1,1,2} \overset{1}{g}_{4,2,3}-\overset{1}{g}_{2,1,2} \overset{1}{g}_{4,1,3}\right)}{\overset{1}{g}_{2,1,2} \left(\overset{1}{g}_{1,1,2}^2+\overset{1}{g}_{2,1,2}^2\right)} \\
 0 & 0 & 0 & \frac{\overset{1}{g}_{4,1,2} \overset{1}{g}_{1,1,2}^2-\overset{1}{g}_{3,1,2} \overset{1}{g}_{4,2,3} \overset{1}{g}_{1,1,2}+\overset{1}{g}_{2,1,2}^2 \overset{1}{g}_{4,1,2}+\overset{1}{g}_{2,1,2} \overset{1}{g}_{3,1,2} \overset{1}{g}_{4,1,3}}{\overset{1}{g}_{1,1,2} \overset{1}{g}_{2,1,2}} \\
 -\frac{\overset{1}{g}_{2,1,2}}{\overset{1}{g}_{1,1,2}} & 1 & 0 & 0 
 \end{bmatrix}.
\end{equation}
Here we see that the condition $\l\neq 0$ is crucial. Commutation relations become
\begin{equation}
\label{eq:commtransf}
    \begin{aligned}\relax
      [\tilde{T}^1, \tilde{T}^2]&=\frac{\overset{1}{g}_{1,1,2} \overset{1}{g}_{3,1,2} \left(\overset{1}{g}_{1,1,2} \overset{1}{g}_{4,1,3}+\overset{1}{g}_{2,1,2} \overset{1}{g}_{4,2,3}\right) \tilde{T}^3}{\overset{1}{g}_{2,1,2} \left(\overset{1}{g}_{4,1,2} \overset{1}{g}_{1,1,2}^2-\overset{1}{g}_{3,1,2} \overset{1}{g}_{4,2,3} \overset{1}{g}_{1,1,2}+\overset{1}{g}_{2,1,2} \left(\overset{1}{g}_{2,1,2} \overset{1}{g}_{4,1,2}+\overset{1}{g}_{3,1,2} \overset{1}{g}_{4,1,3}\right)\right)}, \\
      [\tilde{T}^2, \tilde{T}^4]&= \left(\frac{\overset{1}{g}_{2,1,2}^2}{\overset{1}{g}_{1,1,2}}+\overset{1}{g}_{1,1,2}\right) \tilde{T}^2+\tilde{T}^3\\
      [\tilde{T}^3, \tilde{T}^4]&= \left(\frac{\overset{1}{g}_{2,1,2}^2}{\overset{1}{g}_{1,1,2}}+\overset{1}{g}_{1,1,2}\right) \tilde{T}^3.
    \end{aligned}
\end{equation}
We see that the maximal ideal is $\mc{I}=\frg_{3,1}$, and following the standard steps of the  classification of Mubarakzyanov we find $\frg_{4,8}$ with $\b=0$. 

It is important to note, that the first commutation relation as written above can be singular even if $\l\neq 0$. Some additional work shows, that although the transformation matrix is different, the actual result is the same.

\textbf{2.} The class changes if the maximal nilpotent ideal becomes abelian, $\mc{I}=3\frg_1$, that is $[\tT^1,\tT^2]=0$ in \eqref{eq:commtransf}, while $\l\neq 0$. This requires
\begin{equation}
    \left(\overset{1}{g}_{1,1,2} \overset{1}{g}_{4,1,3}+\overset{1}{g}_{2,1,2} \overset{1}{g}_{4,2,3}\right)=0,
\end{equation}
and the classes are determined by canonical forms of the matrix $M_\a{}^\b$. It has only two non vanishing eigenvalues and the algebras are decomposable. We find that if $\overset{1}{g}_{3,1,2}\neq 0 $ or $\overset{1}{g}_{4,1,2}\neq 0$ it can be brought to the form
\begin{equation}
    M =
        \begin{bmatrix}
            0 & 0 & 0 \\
            0 & 1 & 1\\
            0 & 0 & 1
        \end{bmatrix},
\end{equation}
corresponding to $\frg_{3,2}\oplus \frg_1$.

\textbf{3.} If $\l \neq 0$ and both $\overset{1}{g}_{3,1,2}= 0 $ and $\overset{1}{g}_{4,1,2}=0$ the matrix can be brought to the diagonal form  $M=\mathrm{diag}[0,1,1]$ and the corresponding algebra is $\frg_{3,3}\oplus \frg_1$.

\textbf{4.} If $\l=0$ the algebra is nilpotent and one finds only two cases: non-decomposable $\frg_{4,1}$ and decomposable $\frg_{3,1}\oplus \frg_1$.

Certainly one is always able to use abelian $4\frg_1$ to construct a Manin triple, hence, we do not list this possibility.

Some possible duals posses Killing forms with considerably complicated eigenvalues. In this case we use the \texttt{Reduce} command to find all possible ways to make these eigenvalues vanish or become equal. Majority of the solutions correspond to both eigenvalues vanish and/or involve the imaginary unit.

\section{Conclusions}

In this work we present a coarse classification of 8d Manin triples based on the classification of 4d real Lie algebras by Mubarakzyanov. Each Manin triple $(\frg,\tfrg,\h)$ include two algebras corresponding to (hidden) isometries of Poisson--Lie T-dual string backgrounds. The full list of all such triples is provided in Section \ref{sec:list}, where the dual algebras are not in the standard form. In this sense the classification is coarse: we determine class of the dual algebra and value of it parameter (is any). The latter is expressed in terms of invariant quantities such as eigenvalues of the Killing form and, in the case of abelian maximal nilpotent ideal, of the matrix $M_\a{}^\b$ defined by $[\tT^\a,\tT^4]=M_\b{}^\a \tT^\b $. We refine this classification by marking whether Manin triple $(\frg, \tfrg)$ in each class can by obtained by a Yang--Baxter deformation of $(\frg,4\frg_1)$. In other words, we search for such constant $r^{ab}=-r^{ba}$, that the following holds
\begin{equation}
    \begin{aligned}
            \tf_c{}^{ab} &= 2 r^{d[a}f_{cd}{}^{b]}, \\
            r^{[a_1|b_1}r^{a2|b2|}f_{b1b2}{}^{a3]}&=0.
    \end{aligned}
\end{equation}

During preparation of this text we have found the paper \cite{Hlavaty:2025zfh} appeared recently, that presents conceptually the same classification of 8d Manin triple. The difference between these results and the results presented here is that we are using invariants to distinguish between classes inside each set of dual solutions. This allows to derive values of parameters for dual algebras in the form, that does not depend on the choice of basis. All found dualities between $d=4$ indecomposable real Lie algebras we present diagrammatically in Fig. \ref{fig:duals}. We also one one Poisson--Lie triality, that includes $\frg_{4,7}$, $\frg_{4,8}$ with $\b=-1/2$ and $\frg_{4,2}$ with $\b=-2$. Also the version currently available on-line does not contain the refinement by Yang-Baxter deformations. We find that both results presented here and in  \cite{Hlavaty:2025zfh} complement each other well.

It would be interesting to continue this work in several directions. One is to refine the classification, that is to list all possible explicit commutation relations for dual algebras in each class in the simplest form, given the initial algebra is in the standard form. For 6d Manin triples such a classification has been presented in \cite{Snobl:2002kq}. Certainly one is interested in geometric realization of the dualities found here. Of particular interest are dualities between 3+1 and 4-dimensional backgrounds. For examples, one may start with the Bianchi II cosmological solution endowed with a pure gauge $B$-field and ask whether it is possible to PL-dualize this background into a 4-dimensional space (plus time).

\section*{Acknowledgments}
The authors would like to thank Elena Lanina and Viacheslav Krivorol for valuable comments. This work has been supported by Russian Science Foundation grant RSCF-20-72-10144.

\appendix

\section{List of classes}
\label{sec:list}

\noindent $\frg_{4,1}$

\begin{enumerate}
    \item Solution 1: $\rk=1$ ($\l = \overset{1}{g}_{1,1,2}^2+\overset{1}{g}_{2,1,2}^2$):
        \begin{enumerate}
            \item $\l\neq 0$, $\mc{I}=\frg_{3,1}$: $\tfrg_{4,1}^1 \simeq\frg_{4,8}(\beta=0)$;
            \item $\l \neq 0$, $\mc{I}=3\frg_{1}$: $\tfrg_{4,1}^1 \simeq \frg_{3,2}\oplus \frg_1$, [YB]
             \item $\l\neq 0$, $\mc{I}=3\frg_1$, decomp: $\tfrg_{4,1}^1 \simeq \frg_{3,3}\oplus \frg_1$, [YB]
            \item $\l=0$, $\mc{I}=\frg_{3,1}$, Nil:  $\tfrg_{4,1}^1 \simeq \frg_{4,1}$, [YB]
            \item $\l=0$, $\mc{I}=\frg_{3,1}$, Nil, decomp:  $\tfrg_{4,1}^1 \simeq \frg_{3,1}\oplus \frg_1$, [YB]
        \end{enumerate}
    \item Solution 2: $\rk=1$,  $\mc{I}=3\frg_1$ \\
    $\l = \overset{2}{g}_{2,1,2}^2-\overset{2}{g}_{3,1,3} \overset{2}{g}_{2,1,2}+\overset{2}{g}_{3,1,3}^2+\overset{2}{g}_{3,1,4} \overset{2}{g}_{4,1,3}$),\\
    $\m_1 = -\overset{2}{g}_{2,1,2}$, $\m_{2,3}=\dfrac{1}{2} \left(\m_1\pm\sqrt{4\l-3 \m_1^2}\right)$
        \begin{enumerate}
            \item $4\l-3 \m_1^2\geq0$, $\m_1\neq 0$: $\tfrg_{4,1}^2 \simeq\frg_{4,5} $ with $\b = \fr{1}{2}\left[1+ \fr{\sqrt{4\l-3 \m_1^2}}{\m_1^2}\right]$, $\b+\g=1$
            \item $4\l-3 \m_1^2<0$, $\m_1\neq 0$ : $\tfrg_{4,1}^2 \simeq\frg_{4,6} $ with $\a= \fr{2 \m_1^2}{\sqrt{3\m_1^2-4\l}}$, $\b=\fr12\a$,
            \item $4\l-3 \m_1^2=0$, $\m_1\neq 0$: $\tfrg_{4,1}^2 \simeq\frg_{4,2} $ with $\b=2$
            \item $\m_1 = 0$, $\l\neq 0$: $\tfrg_{4,1}^2\simeq \frg_{3,4}(h=-1)\oplus \frg_1 $,
            \item $\m_{2} = 0$, $\l\neq 0$: $\tfrg_{4,1}^2\simeq \frg_{3,2}\oplus \frg_1 $, [YB]
             \item $\m_{2} = 0$, $\l\neq 0$: $\tfrg_{4,1}^2\simeq \frg_{3,3}\oplus \frg_1 $, [YB]
            \item $\l= 0$, $\m_1 = 0$, Nil, indecomp.: $\tfrg_{4,1}^2 \simeq\frg_{4,1} $, [YB]
            \item $\l= 0$, $\m_1 = 0$, Nil, decomp.: $\tfrg_{4,1}^2 \simeq\frg_{3,1}\oplus \frg_1 $, [YB]
        \end{enumerate}
    \item Solution 3: $\rk=1$, $\l = 2 \overset{3}{g}_{1,1,2}^2$
        \begin{enumerate}
            \item $\l\neq0$, $\mc{I} = \frg_{3,1}$: $\tfrg_{4,1}^3
             \simeq\frg_{4,8}(\beta=0)$;
            \item $\l\neq 0$, $\mc{I} = 3\frg_{1}$: $\tfrg_{4,1}^3 \simeq \frg_{3,2}\oplus \frg_1  $
            \item $\l\neq 0$, $\mc{I} = 3\frg_{1}$: $\tfrg_{4,1}^3 \simeq \frg_{3,3}\oplus \frg_1  $
            \item $\l=0$, $\mc{I} = \frg_{3,1}$, Nil: $\tfrg_{4,1}^3 \simeq\frg_{4,1}$, [YB]
            \item $\l=0$, $\mc{I} = \frg_{3,1}$, Nil, decomp: $\tfrg_{4,1}^3 \simeq\frg_{3,1}\oplus \frg_{1}$, [YB]
        \end{enumerate}
     \item Solution 4: $\rk = 1$, $\mc{I} = 3\frg_1$:\\
     $\l = 2 \left(\overset{4}{g}_{3,1,3}^2+\overset{4}{g}_{3,1,4} \overset{4}{g}_{4,1,3}\right)$, $\m_{1,2}= \pm\sqrt{1/2\l}$
        \begin{enumerate}
            \item $\l > 0$: $\tfrg_{4,1}^4\simeq \frg_{3,4}(h=-1)\oplus \frg_1$
            \item $\l < 0$: $\tfrg_{4,1}^4\simeq \frg_{3,5}(\b=0)\oplus \frg_1$;
            \item $\l=0$, Nil: $\tfrg_{4,1}^4 \simeq  \frg_{4,1}  $, [YB]
            \item $\l=0$, Nil, decomp: $\tfrg_{4,1}^4 \simeq  \frg_{3,1}\oplus \frg_{1}  $, [YB]
        \end{enumerate}
     \item Solution 5: $\rk = 3$, \\
     $\l_1=2 \overset{5}{g}_{2,1,3}^2$, $\l_{2,3} = \overset{5}{g}_{2,1,3} \left(\overset{5}{g}_{3,1,2}\pm\sqrt{4 \overset{5}{g}_{2,1,3}^2+\overset{5}{g}_{3,1,2}^2}\right)$
        \begin{enumerate}
            \item $\l_{1,2}>0$, $\l_3 <0$, semi-simple, decomp: $\tfrg_{4,1}^5\simeq\frg_{3,6}\oplus \frg_1$;
            \item $\l_1=\l_2=\l_3=0$, Nil: $\tfrg_{4,1}^5\simeq  \frg_{4,1}  $, [YB]
             \item $\l_1=\l_2=\l_3=0$, Nil, decomp: $\tfrg_{4,1}^5\simeq\frg_{3,1}\oplus \frg_1$, [YB]
        \end{enumerate}
    \item Solution 6: $\rk=1$, $\mc{I} = 3\frg_1$, \\
       $\l = 2 \left(\overset{6}{g}_{2,1,2}^2+\overset{6}{g}_{3,1,4} \overset{6}{g}_{4,1,3}\right)$,\\
       $\m_1 = -\overset{6}{g}_{2,1,2} $, $\m_{2,3} = \fr12 \left( \m_1 \pm \sqrt{2 \l - 3 \m_1^2}\right)$
        \begin{enumerate}
            \item $2 \l - 3 \m_1^2 >0$, $\m_1 \neq 0$:  $\tfrg_{4,1}^6\simeq\frg_{4,5}$ with $\b=\frac{1}{2}+\sqrt{2 \l - 3 \m_1^2}$, $\b+\g=1$.,
            \item  $2 \l - 3 \m_1^2 <0$, $\m_1 \neq 0$:  $\tfrg_{4,1} ^6\simeq\frg_{4,6}$ with $\a=\frac{2 \overset{6}{g}_{2,1,2}}{\sqrt{3 \m_1^2-2 \lambda}}$, $\b=\fr12\a$, ($\l=0$ allowed)
            \item $\l=0$, $\m_1\neq 0$: $\tfrg_{4,1}^6\simeq \frg_{4,6}$ with $\a = -\fr2{\sqrt{3}}$, $\b =\fr12\a$
            \item $\m_1=0$, $\l > 0$: $\tfrg_{4,1}^6\simeq \frg_{3,4}(h=-1)\oplus \frg_1 $
            \item $\m_1=0$, $\l < 0$: $\tfrg_{4,1}^6\simeq \frg_{3,5}(\b=0)\oplus \frg_1 $
            \item $\m_1=0$, $\l = 0$, Nil: $\tfrg_{4,1}^6\simeq \frg_{4,1}$, [YB]
            \item $\m_2=0$, $\m_{1,3}\neq 0$, $\l = 0$: $\tfrg_{4,1}^6\simeq \frg_{3,2}\oplus \frg_1$
            \item $\m_2=0$, $\m_{1,3}\neq 0$, $\l = 0$: $\tfrg_{4,1}^6\simeq \frg_{3,3}\oplus \frg_1$
        \end{enumerate}
     \item Solution 7: $\rk = 1$, $\l = 2 \overset{7}{g}_{2,1,2}^2$, 
        \begin{enumerate}
            \item $\l\neq 0$, $\mc{I} = \frg_{3,1}$: $\tfrg_{4,1}^7 \simeq\frg_{4,8}(h=0)$
            \item $\l = 0$, Nil: $\tfrg_{4,1}^7 \simeq\frg_{4,1}$, [YB]
            \item $\l = 0$, Nil, decomp: $\tfrg_{4,1}^7 \simeq\frg_{3,1}\oplus \frg_{1}$, [YB]
            \item $\l\neq 0$, $\mc{I} = 3\frg_{1}$: $\tfrg_{4,1}^7 \simeq\frg_{3,2}\oplus \frg_{1}$, [YB]
            \item $\l\neq 0$, $\mc{I} = 3\frg_{1}$: $\tfrg_{4,1}^7 \simeq\frg_{3,3}\oplus \frg_{1}$, [YB]
        \end{enumerate}
    \item Solution 8, $\rk = 1$, $\mc{I} =  3\frg_1$: \\
        $\l=2 \overset{8}{g}_{3,1,4} \overset{8}{g}_{4,1,3}$, 
        $\m_1 = 0$, $\m_{2,3}= \pm \sqrt{\fr\l 2}$
        \begin{enumerate}
            \item $\l > 0$: $\tfrg_{4,1}^8\simeq \frg_{3,4}\oplus \frg_1$ with $\a=-1$
            \item $\l < 0$: $\tfrg_{4,1}^8\simeq \frg_{3,5}\oplus \frg_1$ with $\b=0$
            \item $\l = 0$, Nil: $\tfrg_{4,1}^8\simeq \frg_{4,1}$, [YB]
            \item $\l = 0$, Nil, decomp: $\tfrg_{4,1}^8\simeq \frg_{3,1}\oplus \frg_1$, [YB]
        \end{enumerate}
     \item Solution 9: $\rk = 0$, nilpotent
        \begin{enumerate}
            \item  Indecomposable: $\tfrg_{4,1}^9\simeq \frg_{4,1} $, [YB]
            \item  Decomposable: : $\tfrg_{4,1}^9\simeq \frg_{3,1}\oplus\frg_1 $, [YB]
        \end{enumerate}
\end{enumerate}

\noindent $\frg_{4,2}$ 

\begin{enumerate}
    \item Solution 1, $\rk = 1$, $\mc{I} = 3\frg_1$: \\
        $\l = \left(\beta ^2+2\right) \overset{1}{g}_{4,2,4}^2$, \\
        $\m_1 = \b \overset{1}{g}_{4,2,4} $, $\m_{2,3}= \pm \overset{1}{g}_{4,2,4}$
        \begin{enumerate}
             \item $\l\neq 0$, ($-1 < \b \leq 1$): $\tfrg_{4,2}^1\simeq\frg_{4,5}$ with $(\g'=-1, \b'=\b)$, [YB]
             \item $\l\neq 0$,  ($1<\b $) : $\tfrg_{4,2}^1\simeq\frg_{4,5}$ with $(\g'+\b'=0)$ ,  $(\b'=1/\b)$, [YB] 
             \item $\l\neq 0$,  ($\b<-1 $) : $\tfrg_{4,2}^1\simeq\frg_{4,5}$ with $(\g'+\b'=0)$, $(\g'=1/\b)$, [YB]
            \item $\l\neq 0$,  ($\beta=-1$, $\overset{1}{g}_{4,1,2}\overset{1}{g}_{4,2,3} = 0$) :$\tfrg_{4,2}^1\simeq\frg_{4,5}$ with $\g'=-1,\beta' = -1$, [YB]
            \item$\l\neq 0$,   ($\beta=-1$, $\overset{1}{g}_{4,2,4}\overset{1}{g}_{4,1,2}\overset{1}{g}_{4,2,3} \neq 0$) :$\tfrg_{4,2}^1\simeq\frg_{4,2}$ with $\beta_{new} = -1$, [YB]
            \item $\l=0$: $\tfrg_{4,2}^1\simeq \frg_{3,1}\oplus \frg_1$, [YB]         
        \end{enumerate}
    \item Solution 2, $\rk = 0$:
        \begin{enumerate}
             \item Nil, decomp: $\tfrg_{4,2}^2\simeq \frg_{3,1}\oplus \frg_1$, [YB]
        \end{enumerate}
    \item Solution 3, $\rk =1$, $\mc{I}=3\frg_1$,\\
    $\l=\left(\beta ^2+2\right) \overset{3}{g}_{2,1,2}^2$, \\
    $\m_{1,2}=-\overset{3}{g}_{2,1,2}$, $\m_3=\b \overset{3}{g}_{2,1,2}$
        \begin{enumerate}
             \item $\l \neq0$, $\m_{1,2} \neq \m_3$: $\tfrg_{4,2}^3\simeq\frg_{4,2}$ with $\beta_{new} = -\b_{old}$, [YB]
             \item $\l \neq0$, $\m_{1,2} = \m_3$ : $\tfrg_{4,2}^3\simeq\frg_{4,4}$, [YB]
             \item $\l \neq0$, $\m_3=0$ : $\tfrg_{4,2}^3\simeq\frg_{3,2}\oplus \frg_1$, [YB]
             \item $\l = 0$: $\tfrg_{4,2}^3\simeq\frg_{3,1}\oplus \frg_1$, [YB]   
        \end{enumerate}
    \item Special solution 1 ($\beta = -2$)
        \begin{enumerate}
            \item ($\overset{1}{g}_{4,1,3}\neq 0$ and $\overset{1}{g}_{4,2,4}\neq 0$) $I=\frg_{3,1}$: $\tfrg_{4,2}^1{}'=\frg_{4,8}$ with $h=-1/2$, [YB]
        \end{enumerate}
    \item Special solution 2 ($\beta = -2$)
        \begin{enumerate}
            \item ($\overset{1}{g}_{4,1,4}\neq 0$ and $\overset{1}{g}_{4,2,3}\neq 0$) $I=\frg_{3,1}$: $\tfrg_{4,2}^1{}'=\frg_{4,7}$, [YB]
        \end{enumerate}
\end{enumerate}

\noindent $\frg_{4,3}$

\begin{enumerate}
    \item Solution 1, $\rk = 2$ \\
        \begin{equation*}
            \begin{aligned}
                \l_{1,2}=&\ \frac{1}{2} \Big(\overset{1}{g}_{1,1,2}^2+2 \overset{1}{g}_{1,1,3}^2+\overset{1}{g}_{1,1,4}^2+\overset{1}{g}_{3,2,3}^2+\overset{1}{g}_{4,1,4}^2+2 \overset{1}{g}_{1,1,3} \overset{1}{g}_{3,2,3}\Big) \\ &\pm \fr12 \Bigg[\overset{1}{g}_{1,1,2}^4+2 \left(2 \overset{1}{g}_{1,1,3}^2+2 \overset{1}{g}_{3,2,3} \overset{1}{g}_{1,1,3}+\overset{1}{g}_{1,1,4}^2+\overset{1}{g}_{3,2,3}^2+\overset{1}{g}_{4,1,4}^2\right) \overset{1}{g}_{1,1,2}^2\\
                &+\left(\overset{1}{g}_{1,1,4}^2+\overset{1}{g}_{4,1,4}^2-\overset{1}{g}_{3,2,3} \left(2 \overset{1}{g}_{1,1,3}+\overset{1}{g}_{3,2,3}\right)\right){}^2\Bigg]^{\fr12}
            \end{aligned}
        \end{equation*}
        \\
        either $\l_1$ or $\l_2$ is non-zero
        \begin{enumerate}
            \item $\l_{1,2}\neq 0$: $\tfrg_{4,3}^1\simeq2\frg_{2}$
            \item $\l_2 = 0$: $\tfrg_{4,3}^1\simeq\frg_{4,3}$
        \end{enumerate}
    \item Solution 2, $\rk = 2$, \\
        $\l_1=\overset{2}{g}_{4,1,4}^2\neq 0$, $\l_2 =\overset{2}{g}_{3,2,3}^2$:
        \begin{enumerate}
             \item $\l_{1,2} \neq 0$ : $\tfrg_{4,3}^2\simeq2\frg_{2}$ rank $\k$=2; 
             \item  $\l_2= 0$ :
             $\tfrg_{4,3}^2\simeq\frg_{4,3}$ rank $\k$=1, [YB]
        \end{enumerate}
    \item Solution 3, $\rk =1 $, \\
    $\l=2 \overset{3}{g}_{1,1,2}^2$:
        \begin{enumerate}
             \item $\l\neq 0$, $\mc{I} = \frg_{3,1}$: $\frg_{4,8}(\beta=-1)$,
              \item $\l\neq 0$, $\mc{I}=3\frg_1$ :
             $\tfrg_{4,3}^3\simeq \frg_{3,4}(h=-1)\oplus \frg_1$
             \item $\l= 0$, Nil, decomp : $\tfrg_{4,3}^3\simeq\frg_{3,1}\oplus\frg_1$, [YB]       
        \end{enumerate}
     \item Solution 4, $\rk =1$, $\mc{I} = 3\frg_1$\\
      $\l= \overset{4}{g}_{1,1,2}^2+\overset{4}{g}_{3,2,3}^2$, \\
      $\m_1=0$, $\m_2=\overset{4}{g}_{1,1,2} $, $\m_3=-\overset{4}{g}_{3,2,3}$:
        \begin{enumerate}
             \item $\l\neq 0$, $\m_{2,3} \neq 0$: $\tfrg_{4,3}^4\simeq \frg_{3,4}\oplus \frg_1$, $h = \fr{\m_3}{\m_2}$
             \item $\l\neq 0$, $\m_{2}=0$ or $\m_3=0$: $\tfrg_{4,3}^4\simeq \frg_{4,3}$, [YB]
             \item  $\l\neq 0$, $\m_{2}=0$ or $\m_3=0$, decomp:  $\frg_{2}\oplus 2 \frg_1$, [YB]
             \item $\l= 0$, Nil, decomp. : $\tfrg_{4,3}^4\simeq\frg_{3,1}\oplus\frg_1$, [YB]
        \end{enumerate}
     \item Solution 5, $\rk=2$, \\
     $\l_{1,2}=\frac{1}{2} \left(\overset{5}{g}_{1,1,2}^2+2 \overset{5}{g}_{1,1,3}^2\pm\overset{5}{g}_{1,1,2}\sqrt{\overset{5}{g}_{1,1,2}^2+4 \overset{5}{g}_{1,1,3}^2 }\right)$: ($\overset{5}{g}_{1,1,3}+\overset{5}{g}_{1,1,2}\neq0$)
        \begin{enumerate}
             \item $\l_{1,2}\neq 0$: $\tfrg_{4,3}^5\simeq 2\frg_{2}$,
             \item $\l_2=0$, indecomp.: $\tfrg_{4,3}^5\simeq \frg_{4,3}$
              \item $\l_2=0$, decomp.: $\tfrg_{4,3}^5\simeq\frg_{2}\oplus2\frg_1$, [YB]
        \end{enumerate}
     \item Solution 6, $\rk =2$\\
        \begin{equation*}
            \begin{aligned}
                  \l_{1,2}=&\ \frac{1}{2} \Big(\overset{6}{g}_{1,1,2}^2+2 \overset{6}{g}_{1,1,3}^2+\overset{6}{g}_{1,1,4}^2+\overset{6}{g}_{3,2,3}^2+2 \overset{6}{g}_{1,1,3} \overset{6}{g}_{3,2,3} \\
                  &\pm\fr12 \Bigg[\overset{6}{g}_{1,1,2}^4+2 \left(2 \overset{6}{g}_{1,1,3}^2+2 \overset{6}{g}_{3,2,3} \overset{6}{g}_{1,1,3}+\overset{6}{g}_{1,1,4}^2+\overset{6}{g}_{3,2,3}^2\right) \overset{6}{g}_{1,1,2}^2\\
                  &+\left(\overset{6}{g}_{1,1,4}^2-\overset{6}{g}_{3,2,3} \left(2 \overset{6}{g}_{1,1,3}+\overset{6}{g}_{3,2,3}\right)\right){}^2\Bigg]^{\fr12}
            \end{aligned}
        \end{equation*}   
     either $\l_1$ or $\l_2$ is non-zero:
        \begin{enumerate}
             \item   ($ \overset{6}{g}_{1,1,3}+\overset{6}{g}_{3,2,3}\neq 0$ ): $\tfrg_{4,3}^6\simeq 2\frg_{2}$, rank $\k$=2
             \item   ($ \overset{6}{g}_{1,1,3}+\overset{6}{g}_{3,2,3}= 0$ ): $\tfrg_{4,3}^6\simeq \frg_{4,3}$,  rank $\k$=1, [YB]
        \end{enumerate}
     \item Solution 7, $\rk =2$, \\
     $\l_{1,2}=\pm\frac{1}{2} \left(\sqrt{5}+3\right) \overset{7}{g}_{1,1,3}^2$:
        \begin{enumerate}
             \item $\l_{1,2}\neq 0$: $\tfrg_{4,3}^7\simeq 2\frg_{2}$,
             \item $\l_{1,2}=0$, Nil, decomp.: $\tfrg_{4,3}^7\simeq\frg_{3,1}\oplus\frg_1$ 
        \end{enumerate}
\end{enumerate}

\noindent  $\frg_{4,4}$ 
\begin{enumerate}
    \item Solution 1: $\rk = 1$, $\mc{I} = 3\frg_1$, \\
    $\l = 3 \overset{1}{g}_{4,1,4}^2$, $\m_1 = -\overset{1}{g}_{4,1,4}$, $\m_{2,3} = \overset{1}{g}_{4,1,4}$:
        \begin{enumerate}
            \item $\l \neq 0$:  $\tfrg_{4,4}^1\simeq \frg_{4,2}$ with $\alpha=-1 $, [YB] 
            \item $\l = 0$ Nil, decomp.: $\tfrg_{4,4}^1\simeq \frg_{3,1}\oplus \frg_1$, [YB]
        \end{enumerate}
    \item Solution 2: $\rk = 0$, Nil, decomp:
        \begin{enumerate}
         \item  $\tfrg_{4,4}^2\simeq\frg_{3,1}\oplus\frg_1$, [YB]
        \end{enumerate}
\end{enumerate}

\noindent $\frg_{4,5}$ 

\begin{enumerate}
    \item Solution 1: $\rk = 1$, $\mc{I} = 3\frg_1$, \\
    $\l = {\overset{1}{g}_{4,3,4}^2 \g^{-2}\left(\beta ^2+\gamma ^2+1\right)}$,\\
    $\m_1=\frac{\overset{1}{g}_{4,3,4}}{\gamma }$, $\m_2 =\frac{\beta  \overset{1}{g}_{4,3,4}}{\gamma }$, $\m_3 = -\overset{1}{g}_{4,3,4}$
        \begin{enumerate}
            \item $\l\neq 0$, $\m_{1,2,3}\neq 0$:  $\tfrg_ {4,5}^1\simeq \frg_{4,5}$ with $ (\beta_{new}=\beta, \gamma_{new}=-\gamma)$, [YB]
            \item $\l\neq 0$, $\m_2 = \m_3 \neq \m_1$: $\tfrg_{4,5}^1\simeq \frg_{4,2}$ with           $(\beta_{4,2}=1/\beta)$, [YB]
            \item $\l\neq 0$, $\m_1 = \m_3$: $\tfrg_ {4,5}^1\simeq \frg_{4,2}$ with           $(\beta_{4,2}=\beta)$, [YB]
            \item $\l=0$, Nil, decomp.: $\tfrg_{4,5}^1\simeq\frg_{3,1}\oplus\frg_1$, [YB]
        \end{enumerate}
    \item Solution 2, Nil, decomp.: 
        \begin{enumerate}
         \item $\tfrg_{4,5}^2\simeq\frg_{3,1}\oplus\frg_1$, [YB]
        \end{enumerate}
    \item Solution 3, $\rk=1$, $\mc{I}=3\frg_1$, \\
        $\l=\frac{\overset{3}{g}_{4,2,4}^2 \left(\beta ^2+\gamma ^2+1\right)}{\beta ^2}$, \\
        $\m_1=\frac{\overset{3}{g}_{4,2,4}}{\beta }$, $\m_2=-\overset{3}{g}_{4,2,4}$, $\m_3=\frac{\gamma  \overset{3}{g}_{4,2,4}}{\beta }$:
        \begin{enumerate}
            \item $\l\neq 0$: $\tfrg_ {4,5}^1\simeq \frg_{4,5} $ with           $(\beta_{new}=-\beta, \gamma_{new}=\gamma)$, [YB]
            \item $\l \neq 0$, $\m_2 = \m_3 \neq \m_1$: $\tfrg_ {4,5}^3\simeq \frg_{4,2} $ with  $\beta_{4,2}=-\fr1\beta$, [YB]
            \item $\l \neq 0$, $\m_1 = \m_2$: $\tfrg_ {4,5}^3\simeq \frg_{4,2} $ with  $\beta_{4,2}=\g$, [YB]
            \item $\l=0$, Nil, decomp.: $\tfrg_{4,5}^3\simeq\frg_{3,1}\oplus\frg_1$, [YB]
        \end{enumerate}
    \item Solution 4, $\rk=1$, $\mc{I}=3\frg_1$, \\
        $\l=\overset{4}{g}_{4,1,4}^2 \left(\beta ^2+\gamma ^2+1\right)$, \\
        $\m_1=-\overset{4}{g}_{4,1,4}$, $\m_2 = \beta  \overset{4}{g}_{4,1,4}$, $\m_3 = \gamma  \overset{4}{g}_{4,1,4}$
        \begin{enumerate}
            \item  $\l\neq0$: $\tfrg_ {4,5}^4\simeq \frg_{4,5} $ with $          (\beta_{new}=-\gamma, \gamma_{new}=-\beta)$, [YB]
            \item $\l \neq 0$, $\m_1 = \m_2 $: $\tfrg_ {4,5}^4\simeq \frg_{4,2} $ with  $\beta_{4,2}=-\g$, [YB]
            \item $\l \neq 0$, $\m_1 = \m_3 $: $\tfrg_ {4,5}^4\simeq \frg_{4,2} $ with  $\beta_{4,2}=-\b$, [YB]
            \item $\l=0$, Nil, decomp.: $\tfrg_{4,5}^4\simeq\frg_{3,1}\oplus\frg_1$ , [YB]
        \end{enumerate}
    \item Solution 1' ($\b = 1/2+\a$, $\b+\g=1$): $\rk = 1$, $\l = \frac{2 \left(4 \alpha ^2+3\right) \overset{1}{g}_{4,3,4}^2}{(1-2 \alpha )^2}$ 
        \begin{enumerate}
            \item $\l \neq 0$: $\tfrg_ {4,5}^{1'}\simeq \frg_{4,5}$
            \item $\l = 0$: $\tfrg_{4,5}^{1'}\simeq\frg_{4,1}$
        \end{enumerate}
    \item Solution 2' ($\b+\g=-1$): $\rk = 1$, $\l = \frac{2 \left(\beta ^2+\beta +1\right) \overset{2}{g}_{4,2,4}^2}{\beta ^2}$ 
        \begin{enumerate}
            \item $\l \neq 0$: $\tfrg_ {4,5}^{2'}\simeq \frg_{4,8}$ with $\b'=-1-\
            b$, [YB]
            \item $\l = 0$: $\tfrg_{4,5}^{2'}\simeq\frg_{3,1}\oplus \frg_1$, [YB]
        \end{enumerate}
\end{enumerate} 

\noindent $\frg_{4,6}$ 

\begin{enumerate}
    \item Solution 1: $\rk = 1$, $\mc{I} =3 \frg_1$, \\
        $\l = \frac{ \left(\alpha ^2+2 \beta ^2-2\right)}{\alpha ^2}\overset{1}{g}_{4,1,4}^2$, 
        $\m_1 = -\overset{1}{g}_{4,1,4}$, $\m_{2,3}=\frac{(\beta \pm i) }{\alpha }\overset{1}{g}_{4,1,4}$
        \begin{enumerate}
            \item $\m_1 \neq 0$, $\tfrg_ {4,6}^1\simeq \frg_{4,6} $ with $ \alpha_{new}=-\alpha, \beta_{new}=\beta$ , $\l = 0$ is allowed, [YB]
            \item $\m_1 = 0$, Nil, decomp.:  $\tfrg_ {4,6}^1\simeq \frg_{3,1}\oplus \frg_1 $, [YB]
        \end{enumerate}
    \item Solution 2: nilpotent decomposable
        \begin{enumerate}
         \item $\tfrg_{4,6}^2\simeq\frg_{3,1}\oplus\frg_1$, [YB]
        \end{enumerate}
    \item Solution 1' ($\a+2\b = 0$): 
        \begin{enumerate}
            \item $\overset{1}{g}_{4,1,4} \neq 0$ rank $\k$ =1 : $\tfrg_{4,6}^{1'} = \frg_{4,9}$ with $\a'=\fr12 \a$.
        \end{enumerate}
\end{enumerate}

\noindent $\frg_{4,7}$ 

\begin{enumerate}
    \item Solution 1: $\rk = 1$, $\l = \frac{3}{2} \left(\overset{1}{g}_{4,1,4}^2+4 \overset{1}{g}_{4,2,4}^2\right)$, $\overset{1}{g}_{4,2,4} \neq 0$
        \begin{enumerate} 
            \item $\l \neq 0$: $\mc{I} = \frg_{3,1}$, $\tfrg_{4,7}^1\simeq\frg_{4,8}$ with $(\beta=-1/2)$, [YB]
        \end{enumerate}
    \item Solution 2: $\rk = 1$, $\mc{I} = 3\frg_1$\\
        $\l = \frac{3}{2} \overset{2}{g}_{4,1,4}^2$, \\
        $\m_1 = -\overset{2}{g}_{4,1,4}$, $\m_{2,3}=\frac{1}{2} \overset{2}{g}_{4,1,4}$:
        \begin{enumerate}
            \item $\l \neq 0$: $\tfrg_{4,7}^2\simeq\frg_{4,2}$ with $(\beta=-2)$, [YB]
            \item $\l = 0$: Nil, decomp.: $\tfrg_{4,7}^2 \simeq\frg_{3,1} \oplus\frg_1$, [YB]
        \end{enumerate}
    \item Solution 3: $\rk = 1$, $\l = \frac{3}{2} \overset{3}{g}_{4,1,4}^2$
        \begin{enumerate}
            \item $\l\neq 0$: $\mc{I} = \frg_{3,1}$,  $\tfrg_{4,7}^3\simeq\frg_{4,7}$, [YB]
            \item $\l = 0$: $\tfrg_{4,7}^3\simeq\frg_{3,1}\oplus\frg_1$, [YB]
        \end{enumerate}
\end{enumerate}

\noindent$\frg_{4,8}$ 

\begin{enumerate}
    \item Solution 1, $\rk = 1$, $\mc{I} = \frg_{3,1}$\\
    $\l =\frac{2 \left(\beta ^2+\beta +1\right)}{(\beta +1)^2} \left((\beta +1)^2 \overset{1}{g}_{4,2,4}^2+\overset{1}{g}_{2,1,2}^2\right)$, $\overset{1}{g}_{4,2,4}\neq 0$
        \begin{enumerate}
            \item $-1 \leq \b \leq 0$: $\tfrg_{4,8}^1\simeq \frg_{4,8}$ with $\b_{new} =-1-\b$, [YB];
            \item $0 \leq \b \leq 1$: $\tfrg_{4,8}^1\simeq \frg_{4,8}$ with $\b_{new} =-\fr{1}{1+\b}$, [YB]
        \end{enumerate}
    \item Solution 2, $\rk = 1$, $\mc{I} = \frg_{3,1}$\\
    $\l = \frac{\left(\beta ^2+\beta +1\right)}{2 \beta ^2} \left(\beta ^2 \overset{2}{g}_{2,1,2}^2+4 \overset{2}{g}_{4,3,4}^2\right)$, $\overset{2}{g}_{4,3,4}\neq 0$:
        \begin{enumerate}
         \item $-1 \leq \b < -1/2$ : $\tfrg_ {4,8}^2\simeq \frg_{4,8}$ with $\b_{new}= -\fr{1+\b}{\b}$, [YB]
         \item $-1/2 < \beta \leq 1$ : $\tfrg_ {4,8}^2\simeq \frg_{4,8}$ with $\b_{new}= -\fr{\b}{1+\b}$, [YB]
          \item $\beta=-1/2$ : $\tfrg_ {4,8}^2\simeq \frg_{4,7}$, [YB]
        \end{enumerate}
    \item Solution 3: $\rk=1$, \\
    $\l= \frac{2 \left(\beta ^2+\beta +1\right)}{\beta^2} \overset{3}{g}_{2,1,2}^2$
        \begin{enumerate}
            \item $\l \neq 0$, $\b \neq -1$: $\mc{I} = \frg_{3,1}$, $\tfrg_ {4,8}^3\simeq \frg_{4,8}$ with $\beta_{new}=\b$, [YB]
            \item $\l \neq 0$, $\b = -1$: $\mc{I} = 3\frg_{1}$, $\tfrg_ {4,8}^3\simeq \frg_{3,4}\oplus \frg_1$ with $\a=-1$, [YB]
            \item $\l = 0$: Nil, decomp.: $\tfrg_ {4,8}^3\simeq \frg_{3,1}\oplus \frg_1$, [YB]
        \end{enumerate}
    \item Solution 4, $\rk = 1$, $\mc{I} = 3\frg_1$, \\
        $\l = 2 \left(\beta ^2+\beta +1\right) \overset{4}{g}_{2,1,2}^2$, \\
        $\m_1 = -\overset{4}{g}_{2,1,2}$, $\m_2=-\beta  \overset{4}{g}_{2,1,2}$, $\m_3=(\beta +1) \overset{4}{g}_{2,1,2}$:
        \begin{enumerate}
            \item $\l\neq 0$, $\m_{1,2,3}\neq 0$: $\tfrg_ {4,8}^4\simeq \frg_{4,5}$ with $ 
          (\b' = -1-\b, \g' =\b)$ for $\b <-1/2$\\ and vice versa for $\b > -1/2$, [YB]
          \item $\l\neq 0$, $\m_2=\m_3\neq \m_1$ and $\overset{4}{g}_{4,1,3}=0$: $\tfrg_ {4,8}^4\simeq \frg_{4,5}$ with $\b=-1/2$, $\g=-1/2$, [YB]
            \item $\l\neq 0$, $\m_2=\m_3\neq \m_1$ and $\overset{4}{g}_{4,1,3}\neq 0$: $\tfrg_ {4,8}^4\simeq \frg_{4,2}$ with $\b=-2$;
                \item $\l\neq0$, $\m_2\m_3 =0$: $\tfrg_ {4,8}^4\simeq \frg_{3,4}\oplus \frg_1$ with $\a=-1$, [YB]
            \item $\l = 0$, Nil, decomp.: $\tfrg_{4,8}^4 \simeq\frg_{3,1}\oplus\frg_1$, [YB]
        \end{enumerate}
     \item Special solution $2''$ ($\b=0$): $\rk = 1$, $\l = 2 \left(\overset{2}{g}_{2,1,2}^2+\overset{2}{g}_{2,1,4} \overset{2}{g}_{4,1,2}\right)$
        \begin{enumerate}
            \item $\l= 0$: $\tfrg_{4,8}^{2''}=\frg_{4,1}$, [YB]
        \end{enumerate}
    \item Special solution $1'''$ ($\b=-1$): $\rk = 2$
        \begin{enumerate}
            \item  $\tfrg_{4,8}^{1'''}=\frg_{4,10}$, [YB]
        \end{enumerate}
\end{enumerate}

\noindent  $\frg_{4,9}$ with $\alpha\neq0$ 

\begin{enumerate}
    \item Solution 3, $\rk=1$, $\mc{I} = 3\frg_1$,\\
        $\l =\frac{\left(3 \alpha ^2-1\right)}{2 \alpha ^2} \overset{3}{g}_{4,1,4}^2$ \\
        $\m_1=-\overset{3}{g}_{4,1,4}$, $\m_{2,3}=\frac{(\alpha \pm i)}{2 \alpha } \overset{3}{g}_{4,1,4}$
        \begin{enumerate}
            \item $\m_{1,2,3}\neq 0$, $\tfrg_ {4,9}^3\simeq  \frg_{4,6}$ with $\a_{4,6} = 2 \alpha_{4,9}$, $\beta_{4,6} =- \a_{4,9}$, [YB]
            \item $\m_{1,2,3}= 0$, Nil, decomp.: $\tfrg_ {4,9}^3\simeq  \frg_{3,1}\oplus \frg_1$, [YB] 
        \end{enumerate}
    \item Solution 4: $\rk = 1$, $\l = \frac{\left(3 \alpha ^2-1\right)}{2 \alpha ^2} \overset{4}{g}_{4,1,4}^2$
        \begin{enumerate}
         \item $\mc{I}=\frg_{3,1}$ ($\overset{4}{g}_{4,1,4}\neq 0$):  $\tfrg_ {4,9}^4\simeq\frg_{4,9}$ $\a_{new} = \a$, [YB]
         \item $\mc{I}=3\frg_{1}$ ($\overset{4}{g}_{4,1,4}=0$: $\tfrg_{4,9}^4\simeq\frg_{3,1}\oplus\frg_1$, [YB]
        \end{enumerate}
\end{enumerate}

\noindent  $\frg_{4,9}$ with $\alpha=0$ 

\begin{enumerate}
    \item Solution 3, $\rk=1$, $\mc{I} = 3\frg_1$,\\
        $\l =2 \overset{3}{g}_{3,1,3}^2-2 \overset{3}{g}_{3,1,2}^2$ \\
        $\m_{1,2}=-\overset{3}{g}_{3,1,3}+i \overset{3}{g}_{3,1,2}$, $\m_{3}=0$
        \begin{enumerate}
            \item $\tfrg_ {4,9}^3\simeq  \frg_{3,5}\oplus \frg_1$ with $\b = \fr{\Re \m_1}{\Im \m_1}$, [YB]
        \end{enumerate}
\end{enumerate}

\noindent $\frg_{4,10}$ 

\begin{enumerate}
    \item Solution 1, $\rk =2$, \\
        \begin{equation*}
            \begin{aligned}
                 \l_{1,2}=&\ \left(\overset{1}{g}_{1,1,4} \overset{1}{g}_{3,1,4}+\overset{1}{g}_{1,2,4} \overset{1}{g}_{3,2,4}\right){}^{-2}\left(\overset{1}{g}_{1,1,4}^2+\overset{1}{g}_{1,2,4}^2+\overset{1}{g}_{3,1,4}^2+\overset{1}{g}_{3,2,4}^2\right) \times \\
                 &\times \Bigg\{2 \overset{1}{g}_{1,1,4} \left(\overset{1}{g}_{1,1,4} \overset{1}{g}_{3,1,4}+\overset{1}{g}_{1,2,4} \overset{1}{g}_{3,2,4}\right) \overset{1}{g}_{1,1,2} -\overset{1}{g}_{1,1,2}^2 \left(\overset{1}{g}_{1,1,4}^2+\overset{1}{g}_{1,2,4}^2\right) \\
                 &\pm\Bigg(\left(\overset{1}{g}_{1,1,4}^2+\overset{1}{g}_{1,2,4}^2\right){}^2 \overset{1}{g}_{1,1,2}^4-4 \overset{1}{g}_{1,1,4} \left(\overset{1}{g}_{1,1,4}^2+\overset{1}{g}_{1,2,4}^2\right) \left(\overset{1}{g}_{1,1,4} \overset{1}{g}_{3,1,4}+\overset{1}{g}_{1,2,4} \overset{1}{g}_{3,2,4}\right) \overset{1}{g}_{1,1,2}^3\\
                 &+8 \overset{1}{g}_{1,1,4}^2 \left(\overset{1}{g}_{1,1,4} \overset{1}{g}_{3,1,4}+\overset{1}{g}_{1,2,4} \overset{1}{g}_{3,2,4}\right){}^2 \overset{1}{g}_{1,1,2}^2\\
                 &-8 \overset{1}{g}_{1,1,4} \left(\overset{1}{g}_{1,1,4} \overset{1}{g}_{3,1,4}+\overset{1}{g}_{1,2,4} \overset{1}{g}_{3,2,4}\right){}^3 \overset{1}{g}_{1,1,2}+4 \left(\overset{1}{g}_{1,1,4} \overset{1}{g}_{3,1,4}+\overset{1}{g}_{1,2,4} \overset{1}{g}_{3,2,4}\right){}^4\Bigg)^{\fr12}\Bigg\}
            \end{aligned}
        \end{equation*}
   
        \begin{enumerate}
         \item $\l_1 \l_2 \neq 0$, $\rk = 2$: $\tfrg_ {4,10}^1\simeq\frg_{4,10}$, [YB]
         \item $\l_1=0$, $\l_2 \neq 0$, $\mc{I}=3\frg_1$, ($\m_{1,2}=u\pm i v, \m_3=0, v\neq 0$):  $\tfrg_ {4,10}^1\simeq\frg_{3,5}\oplus \frg_1$         with $\b = \frac{u}{v} $
         \item $\l_1=0$, $\l_2 \neq 0$, $\mc{I}=3\frg_1$, ($\m_{1,2}=u, \m_3=0$):
         $\tfrg_ {4,10}^1\simeq\frg_{3,3}\oplus \frg_1$ rank $\k$= 1
        \end{enumerate}
    \item Solution 2: $\rk=3$, $\l_1=2 \left(\overset{2}{g}_{1,1,2}^2+\overset{2}{g}_{2,1,2}^2-\overset{2}{g}_{1,2,3} \overset{2}{g}_{3,1,2}\right)$, \\ $\l_{2,3}=-\overset{2}{g}_{1,2,3} \left(\pm\sqrt{4 \overset{2}{g}_{1,1,2}^2+4 \overset{2}{g}_{2,1,2}^2+\left(\overset{2}{g}_{1,2,3}-\overset{2}{g}_{3,1,2}\right){}^2}+\overset{2}{g}_{1,2,3}+\overset{2}{g}_{3,1,2}\right)$
        \begin{enumerate}
            \item   $\l_1\neq 0$, $\l_{2}=0$, $\l_3 = 0$: $\tfrg_ {4,10}^2\simeq \frg_{4,8}(\beta=-1)$, [YB]
            \item $\l_1=0$, $\l_2=0$, $\l_3\neq0$: $\tfrg_ {4,10}^2\simeq \frg_{3,5}\oplus\frg_{1}$ with $\b=0$
            \item $\l_1<0 $, $\l_2\l_3\neq 0$: $\tfrg_ {4,10}^2\simeq \frg_{3,7}\oplus\frg_{1}$,
            \item $\l_1>0 $, $\l_2\l_3\neq 0$: $\tfrg_ {4,10}^2\simeq \frg_{3,6}\oplus\frg_{1}$,
            \item $\l_1=0$, $\l_2=0$, $\l_3=0$: $\tfrg_ {4,10}^2\simeq \frg_{3,1}\oplus\frg_{1}$, [YB] 
        \end{enumerate}
     \item Solution 3: $\rk = 2$, $\l_1 = -2 \left(\overset{3}{g}_{1,1,2}-\overset{3}{g}_{3,1,4}\right){}^2 \overset{3}{g}_{3,1,4}^{-2}\left(\overset{3}{g}_{1,1,4}^2+\overset{3}{g}_{3,1,4}^2+\overset{3}{g}_{3,2,4}^2\right)$, \\
     $\l_2= 2 \left(\overset{3}{g}_{1,1,4}^2+\overset{3}{g}_{3,1,4}^2+\overset{3}{g}_{3,2,4}^2\right)$ 
        \begin{enumerate}
         \item  $\l_1\neq 0$, $\l_2 \neq 0$:  $\tfrg_ {4,10}^2\simeq\frg_{4,10}$, [YB]
         \item  $\l_1 = 0$:  $\tfrg_ {4,10}^2\simeq\frg_{3,3}\oplus \frg_1$, [YB]
        \end{enumerate}
    \item Solution 6, $\rk = 2$\\
        \begin{equation*}
            \begin{aligned}
                \l_{1,2}&= \overset{6}{g}_{1,1,4}^{-2} \overset{6}{g}_{3,2,4}^{-2}\left(\overset{6}{g}_{1,1,4}^2+\overset{6}{g}_{3,2,4}^2\right) \Bigg(-\overset{6}{g}_{1,1,4}^2 \overset{6}{g}_{2,1,2} \left(\overset{6}{g}_{2,1,2}-2 \overset{6}{g}_{3,2,4}\right) \\
                &\pm\Bigg[\left(\overset{6}{g}_{2,1,2}^2-2 \overset{6}{g}_{3,2,4} \overset{6}{g}_{2,1,2}+2 \overset{6}{g}_{3,2,4}^2\right){}^2 \overset{6}{g}_{1,1,4}^4+8 \overset{6}{g}_{1,2,4}^2 \overset{6}{g}_{3,2,4}^3 \left(\overset{6}{g}_{3,2,4}-\overset{6}{g}_{2,1,2}\right) \overset{6}{g}_{1,1,4}^2+4 \overset{6}{g}_{1,2,4}^4 \overset{6}{g}_{3,2,4}\Bigg]^{\fr12}\Bigg)
            \end{aligned}
        \end{equation*}
        \begin{enumerate}
            \item $\l_1 \l_2 \neq 0$: $\tfrg_ {4,10}^6\simeq\frg_{4,10}$, [YB]
           \item $\l_1=0$, $\l_2 \neq 0$, $\mc{I}=3\frg_1$, ($\m_{1,2}=u\pm i v, \m_3=0, v\neq 0$):  $\tfrg_ {4,10}^1\simeq\frg_{3,5}\oplus \frg_1$         with $\b = \frac{u}{v} $, 
         \item $\l_1=0$, $\l_2 \neq 0$, $\mc{I}=3\frg_1$, ($\m_{1,2}=u, \m_3=0$):
         $\tfrg_ {4,10}^1\simeq\frg_{3,3}\oplus \frg_1$ rank $\k$= 1
        \end{enumerate}
    \item Solution 7, $\rk = 2$, \\
        \begin{equation*}
            \begin{aligned}
                \l_{1,2} =& \overset{7}{g}_{1,1,4}^{-2} \overset{7}{g}_{3,1,4}^{-2}\left(\overset{7}{g}_{1,1,4}^2+\overset{7}{g}_{1,2,4}^2+\overset{7}{g}_{3,1,4}^2\right) \Bigg(2 \overset{7}{g}_{1,1,4}^2 \overset{7}{g}_{3,1,4} \overset{7}{g}_{1,1,2}-\overset{7}{g}_{1,1,2}^2 \left(\overset{7}{g}_{1,1,4}^2+\overset{7}{g}_{1,2,4}^2\right)\\
                &\pm\Bigg[\left(\overset{7}{g}_{1,1,4}^2+\overset{7}{g}_{1,2,4}^2\right){}^2 \overset{7}{g}_{1,1,2}^4-4 \overset{7}{g}_{1,1,4}^2 \left(\overset{7}{g}_{1,1,4}^2+\overset{7}{g}_{1,2,4}^2\right) \overset{7}{g}_{3,1,4} \overset{7}{g}_{1,1,2}^3+8 \overset{7}{g}_{1,1,4}^4 \overset{7}{g}_{3,1,4}^2 \overset{7}{g}_{1,1,2}^2\\
                &-8 \overset{7}{g}_{1,1,4}^4 \overset{7}{g}_{3,1,4}^3 \overset{7}{g}_{1,1,2}+4 \overset{7}{g}_{1,1,4}^4 \overset{7}{g}_{3,1,4}^4\Bigg]^{\fr12}\Bigg)
            \end{aligned}
        \end{equation*}
        \begin{enumerate}
            \item $\l_1 \l_2 \neq 0$: $\tfrg_ {4,10}^6\simeq\frg_{4,10}$, [YB]
           \item $\l_1=0$, $\l_2 \neq 0$, $\mc{I}=3\frg_1$, ($\m_{1,2}=u\pm i v, \m_3=0, v\neq 0$):  $\tfrg_ {4,10}^1\simeq\frg_{3,5}\oplus \frg_1$         with $\b = \frac{u}{v} $, 
         \item $\l_1=0$, $\l_2 \neq 0$, $\mc{I}=3\frg_1$, ($\m_{1,2}=u, \m_3=0$):
         $\tfrg_ {4,10}^1\simeq\frg_{3,3}\oplus \frg_1$ rank $\k$= 1,  
        \end{enumerate}
    \item  Solution 8: $\rk=2$,  $\l_{1,2}=\pm2 \left(\overset{8}{g}_{3,1,4}^2+\overset{8}{g}_{3,2,4}^2\right)$
        \begin{enumerate}
         \item $\l_1\neq0$, $\l_2 \neq 0$:  $\tfrg_ {4,10}^8\simeq\frg_{4,10}$, [YB] 
         \item $\l_1 = 0$, $\l_2  = 0$:  $\tfrg_ {4,10}^8\simeq\frg_{3,1}\oplus \frg_1$, [YB]
        \end{enumerate}
    \item Solution 11: $\rk = 2$ \\
    $\l_1 = -{2\overset{11}{g}_{3,2,4}^{-2} \left(\overset{11}{g}_{2,1,2}-\overset{11}{g}_{3,2,4}\right){}^2 \left(\overset{11}{g}_{1,1,4}^2+\overset{11}{g}_{3,2,4}^2\right)}$,  $ \l_2=2 \left(\overset{11}{g}_{1,1,4}^2+\overset{11}{g}_{3,2,4}^2\right)\neq 0$
        \begin{enumerate}
         \item $\l_1 \neq 0$: $\tfrg_ {4,10}^{11}\simeq\frg_{4,10}$, [YB] 
         \item  $\l_1 = 0 $: $\tfrg_ {4,10}^{11}\simeq\frg_{3,3}\oplus \frg_1$ rank $\k$= 1, [YB] 
        \end{enumerate}
    \item Solution 12: $\rk = 2$ \\
        $\l_1 = -2 \overset{12}{g}_{3,1,4}^{-2} \left(\overset{12}{g}_{1,1,2}-\overset{12}{g}_{3,1,4}\right){}^2 \left(\overset{12}{g}_{1,1,4}^2+\overset{12}{g}_{3,1,4}^2\right)$, $\l_2 = 2 \left(\overset{12}{g}_{1,1,4}^2+\overset{12}{g}_{3,1,4}^2\right)\neq 0$
        \begin{enumerate}
         \item $\l_1 \neq 0$: $\tfrg_ {4,10}^{11}\simeq\frg_{4,10}$, [YB]  
         \item  $\l_1 = 0 $: $\tfrg_ {4,10}^{11}\simeq\frg_{3,3}\oplus \frg_1$ rank $\k$= 1, [YB]  
        \end{enumerate}
    \item Solution 13: $\rk = 2$, $\l_{1,2} = \overset{13}{g}_{3,1,4}^{-2}{\left(\overset{13}{g}_{1,2,4}^2+\overset{13}{g}_{3,1,4}^2\right) \left(-\overset{13}{g}_{2,1,2}^2\pm\sqrt{\overset{13}{g}_{2,1,2}^4+4 \overset{13}{g}_{3,1,4}^4}\right)}\neq 0$
        \begin{enumerate}
            \item $\l_{1,2} \neq 0$: $\tfrg_ {4,10}^{13}\simeq\frg_{4,10}$ 
        \end{enumerate}
    \item Solution 16, $\rk = 2$, \\
        $\l_{1,2}=\overset{16}{g}_{1,1,4}^2-\overset{16}{g}_{1,2,3}^2\pm\sqrt{4 \overset{16}{g}_{1,2,4}^4-8 \overset{16}{g}_{1,1,4} \overset{16}{g}_{1,2,3} \overset{16}{g}_{1,2,4}^2+\left(\overset{16}{g}_{1,1,4}^2+\overset{16}{g}_{1,2,3}^2\right){}^2}$
        \begin{enumerate}
            \item $\l_{1,2} \neq 0$: $\tfrg_ {4,10}^{13}\simeq\frg_{4,10}$, [YB]
            \item $\l_1 =0$, $\mc{I} = 3\frg_1$ ($\m_{1,2}=u\pm i v,\m_3=0,v\neq 0$):  $\tfrg_ {4,10}^{16}\simeq\frg_{3,3}\oplus \frg_1$ with $\b=\fr{u}{v}$ 
        \end{enumerate}
    \item Solution 17: $\rk =2$, $\l_{1,2}= \pm 2 \overset{17}{g}_{3,2,4}^2$
        \begin{enumerate}
        \item $\l_{1,2}\neq 0 $: $\tfrg_ {4,10}^{17}\simeq \frg_{4,10}$, [YB]  
         \item $\l_{1,2}=0$ :  $\tfrg_ {4,10}^{17}\simeq\frg_{3,1}\oplus\frg_1$, [YB] 
        \end{enumerate}
    \item Solution 18: $\rk=2$, $\l_{1,2}= \pm 2 \overset{18}{g}_{3,1,4}^2$
        \begin{enumerate}
         \item $\l_{1,2}\neq0$: $\tfrg_{4,10}^{18}\simeq\frg_{4,10}$, [YB]   
         \item $\l_{1,2}=0$: $\tfrg_{4,10}^{18}\simeq\frg_{3,1}\oplus\frg_1$, [YB]   
        \end{enumerate}
    \item Solution 19: $\rk = 1$, $\l= 2 \left(\overset{19}{g}_{1,1,4}^2+\overset{19}{g}_{3,2,4}^2\right)\neq 0$
        \begin{enumerate}
           \item $\tfrg_ {4,10}^{19}\simeq\frg_{3,3}\oplus\frg_1$  
        \end{enumerate}
    \item Solution 20: $\rk=2$, $\l_1 =2 \overset{20}{g}_{1,1,4}^2$, $\l_2=-2 \overset{20}{g}_{1,2,3}^2$
        \begin{enumerate}
          \item $\l_{1,2}\neq 0$: $\tfrg_ {4,10}^{20}\simeq\frg_{4,10}$, [YB] 
         \item  $\l_1=0$: $\tfrg_ {4,10}^{20}\simeq\frg_{3,5}\oplus \frg_1$ with $\b =0 $ ,
         \item  $\l_2 = 0 $: $\tfrg_ {4,10}^{20}\simeq\frg_{3,3}\oplus \frg_1$ 
        \end{enumerate}
    \item Solution 21: $\rk=2$, $\l_{1,2}=-\overset{21}{g}_{1,2,3}^2\pm\sqrt{\overset{21}{g}_{1,2,3}^4+4 \overset{21}{g}_{1,2,4}^4}$
        \begin{enumerate}
         \item  $\l_{1,2}\neq 0$:   $\tfrg_ {4,10}^{20}\simeq\frg_{4,10}$ 
         \item  $\l_1=0$ : $\tfrg_ {4,10}^{21}\simeq\frg_{3,5}\oplus\frg_1$ with $\b = 0$, [YB]
        \end{enumerate}
\end{enumerate}

\bibliography{bib.bib}
\bibliographystyle{utphys.bst}
\end{document}